\documentclass[prd,twocolumn,a4paper,superscriptaddress,floatfix]{revtex4}
\usepackage{graphicx}
\usepackage{bbm}
\usepackage[all]{xy}
\usepackage{amsmath}
\usepackage{amssymb}
\usepackage{epstopdf}

\def\be{\begin{equation}}
\def\ee{\end{equation}}
\newcommand{\bq}{\begin{eqnarray}}
\newcommand{\eq}{\end{eqnarray}}
\newcommand{\bes}{\begin{subequations}}
\newcommand{\ees}{\end{subequations}}

\def\ben{\begin{eqnarray}}
\def\een{\end{eqnarray}}
\def\ba{\begin{array}}
\def\ea{\end{array}}

\begin{document}
\newcommand{\half}{{\textstyle\frac{1}{2}}}
\allowdisplaybreaks[3]
\def\a{\alpha}
\def\b{\beta}
\def\g{\gamma}\def\G{\Gamma}
\def\d{\delta}\def\D{\Delta}
\def\ep{\epsilon}
\def\et{\eta}
\def\z{\zeta}
\def\t{\theta}\def\T{\Theta}
\def\l{\lambda}\def\L{\Lambda}
\def\m{\mu}
\def\f{\phi}\def\F{\Phi}
\def\n{\nu}
\def\p{\psi}\def\P{\Psi}
\def\r{\rho}
\def\s{\sigma}\def\S{\Sigma}
\def\ta{\tau}
\def\x{\chi}
\def\o{\omega}\def\O{\Omega}
\def\k{\kappa}
\def\pa {\partial}
\def\ov{\over}
\def\br{\\}
\def\ud{\underline}

\newcommand\lsim{\mathrel{\rlap{\lower4pt\hbox{\hskip1pt$\sim$}}
    \raise1pt\hbox{$<$}}}
\newcommand\gsim{\mathrel{\rlap{\lower4pt\hbox{\hskip1pt$\sim$}}
    \raise1pt\hbox{$>$}}}
\newcommand\esim{\mathrel{\rlap{\raise2pt\hbox{\hskip0pt$\sim$}}
    \lower1pt\hbox{$-$}}}
\newcommand{\dpar}[2]{\frac{\partial #1}{\partial #2}}
\newcommand{\sdp}[2]{\frac{\partial ^2 #1}{\partial #2 ^2}}
\newcommand{\dtot}[2]{\frac{d #1}{d #2}}
\newcommand{\sdt}[2]{\frac{d ^2 #1}{d #2 ^2}}    

\title{The cosmological evolution of $p$-brane networks}

\author{L. Sousa}
\email[Electronic address: ]{laragsousa@gmail.com}
\affiliation{Centro de F\'{\i}sica do Porto, Rua do Campo Alegre 687, 4169-007 Porto, Portugal}
\affiliation{Departamento de F\'{\i}sica da Faculdade de Ci\^encias
da Universidade do Porto, Rua do Campo Alegre 687, 4169-007 Porto, Portugal}
\author{P.P. Avelino}
\email[Electronic address: ]{ppavelin@fc.up.pt}
\affiliation{Centro de Astrof\'{\i}sica da Universidade do Porto, Rua das Estrelas, 4150-762 Porto, Portugal}
\affiliation{Departamento de F\'{\i}sica da Faculdade de Ci\^encias
da Universidade do Porto, Rua do Campo Alegre 687, 4169-007 Porto, Portugal}

\begin{abstract}

In this paper we derive, directly from the Nambu-Goto action, the relevant components of the acceleration of cosmological featureless $p$-branes, extending previous analysis based on the field theory equations in the thin-brane limit. The 
component of the acceleration parallel to the velocity is at the core of the velocity-dependent one-scale model for the 
evolution of $p$-brane networks. We use this model to show that, in a decelerating expanding universe in which the $p$-branes are relevant cosmologically, interactions cannot lead to frustration, except for fine-tuned non-relativistic networks with a dimensionless curvature parameter $k \ll 1$. We discuss the implications of our findings for the cosmological evolution of $p$-brane networks.

\end{abstract} 
\maketitle

\section{Introduction}

It is generally accepted that the universe underwent a phase of accelerated expansion in its early history. This paradigm, usually  denominated cosmological inflation \cite{Guth:1980zm,Linde:1981mu}, explains the extreme flatness and homogeneity of the observed universe, as well as the the origin of the large scale structure. In the context of the brane-world realization of string theory, cosmological inflation could be driven by the interaction between p-dimensional D-branes \cite{Dvali:1998pa,Burgess:2001fx,ALexander:2001ks} (which are, along with fundamental strings, the fundamental objects of the theory). Brane inflationary  scenarios typically end with a symmetry breaking phase transition, leading to the production of branes with lower dimensionality. The copious production of $1$-dimensional branes (cosmic strings) is predicted for a large variety of brane inflation models, while higher-dimensional $p$-branes might also be produced \cite{Sarangi:2002yt,Majumdar:2002hy,Jones:2002cv}. Therefore, brane inflation may lead to the production of $p$-branes networks, evolving in higher dimensional backgrounds. 

Cosmic strings and domain walls are the only non-trivial $p$-brane solutions allowed in $3+1$ dimensional backgrounds. The interest in cosmic strings has recently been revived due to the possibility of detecting their cosmological signatures observationally. In particular, they may leave an observable imprint on the B-mode polarization \cite{Seljak:2006hi,Pogosian:2007gi} and small-scale anisotropy \cite{Pogosian:2008am} of the  cosmic microwave background which may be within reach of the Planck mission. Moreover, there is also the prospect of the future detection of the gravitational waves emitted by cosmic strings with the LIGO2 and LISA probes \cite{Polchinski:2007rg} (see also \cite{Avelino:2003nn,Morganson:2009yk,Brandenberger:2010hn} for other examples of possible observational signatures of cosmic strings). In order to make precise observational predictions it is necessary to understand the nature of cosmic strings and their late-time evolution.  While standard cosmic string networks have been extensively studied both using a semi-analytical velocity-dependent one-scale (VOS) model \cite{VOS} and numerical simulations \cite{Allen:1990tv}, significantly less is known about the dynamics of complex string networks with junctions. For example, cosmic superstring networks may have an hierarchy of tensions and junctions \cite{Copeland:2003bj} and, consequently, their observational signatures are expected to be different from those of ordinary cosmic strings \cite{Brandenberger:2008ni,Danos:2009vv}. Still, it is not yet clear whether or not these differences prevent cosmic superstring networks from attaining a linear scaling regime and, therefore, the late-time evolution of these networks is not yet completely understood \cite{Copeland:2005cy,Avgoustidis:2007aa,Pourtsidou:2010gu}.

A $p$-brane network, if frozen in comoving coordinates (or frustrated), is characterized by a negative average pressure. This property is particularly interesting in the case of domain wall networks. There is compelling evidence \cite{Komatsu:2010fb,Amanullah2010} which indicates that the universe is currently undergoing a phase of accelerated expansion, caused by an exotic energy component (dubbed dark energy) accounting for more than two thirds of the energy density of the universe and whose nature is yet unknown. If domain walls were the dominant energy component, and provided their root-mean-square (RMS) velocity was small enough, they could drive a phase of accelerated expansion. For this reason, frustrated domain wall networks have been suggested as a dark energy candidate \cite{Bucher:1998mh}. The conditions under which domain wall networks may frustrate as a result of cosmological evolution were intensively studied in \cite{PinaAvelino:2006ia,Avelino:2006xy,Battye:2006pf,Avelino:2006xf,Avelino:2008ve,Sousa:2009is} using a semi-analytical VOS model and field-theory simulations. These studies indicate that frustration does not result naturally from the evolution of realistic and cosmologically relevant domain wall networks. 

The cosmological dynamics of generic $p$-brane networks has been studied in \cite{Sousa:2011ew} using a VOS model characterizing the evolution of their RMS velocity and characteristic length. As in the case of cosmic string \cite{Martins:1996jp} and domain wall networks \cite{Avelino:2006xf,Avelino:2011ev}, in homogeneous and isotropic universes with a decelerating power law expansion, $p$-brane networks may evolve towards a linear scaling solution in which the characteristic length grows proportionally to the Hubble radius and the RMS velocity remains constant \cite{Sousa:2011ew}. This scale-invariant solution is an attractor of the VOS equations, characterizing the late-time evolution of the networks in a frictionless regime. 

In this paper we derive the evolution equation for the velocity of featureless infinitely-thin $p$-brane networks in homogeneous and isotropic universes with an arbitrary number of spatial dimensions, directly from the Nambu-Goto action. This generalizes previous results obtained using field theory equations in the thin-brane limit, further validating the VOS model for the evolution of $p$-brane networks in $N+1$-dimensional homogeneous and isotropic universes derived in \cite{Sousa:2011ew}. We also consider the effect of a generic interaction mechanism between the $p$-branes and other cosmological components, studying its  potential role in the frustration of the networks.

Throughout the work, we will assume the metric signature $[+,-,...,-]$ and the calculations will be done using units in which $c=1$. The Einstein summation convention will be used when a greek index variable appears twice in a single term, once in an upper (superscript) and once in a lower (subscript) position.

\section{Featureless $p$-branes: equation of state\label{eos-sec}}

Consider a local inertial frame in which the $p$-brane is instantaneously at rest. If the $p$-brane is locally flat, a set of planar orthogonal coordinates $\left(t,x^1,\cdots,x^N\right)$ may be chosen such that $x^1,\cdots,x^p$ parameterize the brane and $x^{p+1},\cdots,x^N$ are perpendicular to it. Since the properties of a featureless $p$-brane do not change along the parallel directions, there is no component of the physical velocity along these directions. Therefore, the components of the energy-momentum tensor $T^{\mu\nu}$ must be invariant with respect to Lorentz boosts in any direction along the brane.

Consider a boost along one of the parallel directions $x^{\tilde i}$ (${\tilde i}=1,\cdots,p$ throughout the paper). The energy-momentum tensor, transforms as
\be
T^{\mu'\nu'}=\Lambda^{\mu'}_{\alpha}\Lambda^{\nu'}_{\beta}T^{\alpha\beta}\,,
\ee
where
\bq
\Lambda^{0'}_0=\Lambda^{\tilde {i'}}_{\tilde i}=\gamma\,, \quad \quad \Lambda^{0'}_{\tilde i}=\Lambda^{\tilde {i'}}_0=\gamma v\,,\\
\Lambda^{l'}_l=1\,,\quad \mbox {for} \quad l \neq {\tilde i}\,,
\eq
and all other components vanish. Here $l=1,\cdots,N$.

Hence 
\bq
T^{0'l'}=\gamma T^{0l}+\gamma v T^{{\tilde i}l}=T^{0l}\,,\\
T^{{\tilde {i'}}l'}=\gamma T^{{\tilde i}l}+\gamma v T^{0l}=T^{{\tilde i}l}\,,\\
\eq
which leads to
\be
T^{0l}=T^{{\tilde i}l}=0\,,
\ee
for $l \neq {\tilde i}$. It is possible to show using similar arguments that $T^{0{\tilde i}}$ must also vanish. Moreover,
\be
T^{0'0'}=\gamma^2 T^{00}+\gamma^2 v^2 T^{{\tilde i}{\tilde i}}=T^{00}\,,
\ee
which gives
\be
T^{{\tilde i}{\tilde i}}=-T^{00}\,.
\ee

If the $p$-brane is maximally symmetric with respect to the $N-p$ perpendicular directions and its energy is localized then Derrick's theorem \cite{1964JMP.....5.1252D} implies that a necessary condition for stability is \cite{Avelino:2010bu} 
\be
\int d^D x T^{ll}=0\,,
\ee
for $l \ge p+1$ (here $D=N-p$ and $d^D x=dx^{p+1} \times \,...\, \times dx^N$). Spherical symmetry with respect to the $D$ perpendicular directions implies that, at the core, we should also have that $T^{lm}=0$, for $l \ge p+1$,  $m \ge p+1$ and $l \neq m$. In most situations of cosmological interest, the $p$-brane thickness is very small compared to its curvature radii and may therefore be neglected. Hence, if the $p$-brane is infinitely thin, the non-vanishing components of the energy-momentum tensor are
\bq	
T^{00}&=&\sigma_p \int d^p x \, \delta^N ({\bf x}-{\bf x}_p)  \,,\\ 
T^{{\tilde i}{\tilde i}}&=&-\sigma_p \int  d^p x \, \delta^N ({\bf x}-{\bf x}_p)\,,
\eq
where $\sigma_p$ is the (constant) $p$-brane mass per unit $p$-dimensional area, ${\bf x}$ is a N-vector whose components are cartesian coordinates, ${\bf x}_p$ represents the $p$-brane profile, and $\delta^N({\bf x})$ is the $N$-dimensional Dirac delta function.

The equation-of-state parameter of a brane gas can be obtained by performing a Lorentz boost to the energy-momentum tensor of a single static flat $p$-brane and, then, averaging over all possible orientations of the brane \cite{kolb}
\be
w=\frac{\bar {\mathcal P}}{\bar \rho}=\frac{1}{N}\left[\left(p+1\right){\bar v}^2-p\right]\,.
\label{eos}
\ee
Here, ${\bar v}$ is the RMS velocity of the branes, ${\bar \rho}=V^{-1}\int \rho dV$ is the average brane density, ${\bar {\mathcal P}}=V^{-1}\int\mathcal{P}dV$ is the average brane pressure and $V$ is a large volume.

Note that Eq. (\ref{eos}) has two important limits. In the relativistic limit, with ${\bar v} \to 1$,  one has $w \to 1/N$ independently of the value of $p$. In the non-relativistic limit (${\bar v} \to 0$) one has $w=-p/N$. Taking into account that the Raychaudhury  in a $N$+1 dimensional FRW universe is given by
\be
\frac{\ddot a}{a} = -\frac{8 \pi G_{N+1}}{N(N-1)} \left( (N-2) \rho_b + N  \mathcal{P}_b\right)\,,
\ee
where $\rho_b$ is the background density, $\mathcal{P}_b$ is the background pressure and $G_{N+1}$ is the $N+1$ dimensional Newton constant, we conclude that only domain walls ($p$-branes with $p=N-1$) or a cosmological constant ($p$-brane with $p=N$) could lead to an accelerated expansion of the universe.

\section{$p$-brane dynamics\label{act-sec}}

In the zero-thickness limit, the world-history in spacetime of a featureless $p$-brane may be represented by
\be
x^{\mu}=x^{\mu}(u^{{\tilde \nu}})\,,
\ee
where $u^{\tilde \nu}$ with ${\tilde \nu}=0,1,...,p$ are the coordinates parameterizing the $(p+1)$-dimensional worldsheet swept by the $p$-brane, $u^0$ is a timelike parameter and $u^{\tilde i}$ are spacelike parameters. The action of the $p$-brane may be derived from the underlying field theory
\be
S=\int d^{N+1}x\sqrt{\left|g\right|}\mathcal{L}\,,
\ee
where $g=\det(g_{\mu\nu})$, $g_{\mu\nu}$ is the metric tensor and $\mathcal{L}$ is the lagrangian of the model. In the vicinity of the $p$-brane, the line element can be written as \cite{Vilenkin}
\be
ds^2={\tilde g}_{{\tilde \mu} {\tilde \nu}}d x^{\tilde \mu} d x^{\tilde \nu} + d {\bf r} \cdot d {\bf r}\,,
\ee
where ${\tilde g}_{{\tilde \mu} {\tilde \nu}}=g_{\alpha \beta} x^{\alpha}_{,{\tilde \mu}}x^{\beta}_{,{\tilde \nu}}$ is the worldsheet metric with $x^{\alpha}_{,{\tilde \mu}}=\partial x^{\alpha}/\partial u_{\tilde \mu}$ and $(d{\bf r}\cdot d{\bf r})^{1/2}$ is the infinitesimal distance to the brane ($p$-dimensional) core. Therefore, the volume element is given by
\be
d^{N+1}x=\sqrt{\left|{\tilde g}\right|}d^{p+1} u d^{D}x\,.
\ee
Since one is assuming that the $p$-brane is thin and featureless, the lagrangian density can only vary along the perpendicular directions and, as consequence, it depends only on the $x^{p+1},\cdots,x^{N}$ coordinates. Integrating the action with respect to these coordinates, one obtains the Nambu-Goto action for infinitely thin $p$-branes
\be
S=-\sigma_p \int d^{p+1}u\sqrt{\left|{\tilde g}\right|}\,,
\label{nambugoto}
\ee
where 
\be
\sigma_p=-\int d^{D}x\mathcal{L} 
\ee
is the (constant) $p$-brane mass per unit $p$-dimensional area.

\section{Equation of motion\label{eom-sec}}

By varying the Nambu-Goto action in Eq. (\ref{nambugoto}) with respect to $x^{\mu}$, one obtains the equations of motion for the dynamic variables $x^{\mu}$
\be
\frac{1}{\sqrt{\left|{\tilde g}\right|}} \left(\sqrt{\left|{\tilde g}\right|} {\tilde g}^{{\tilde \mu} {\tilde \nu}} {x^{\alpha}}_{,\tilde \nu}\right)_{,{\tilde \mu}}+\Gamma^{\alpha}_ {\beta\lambda}{\tilde g}^{{\tilde \mu} {\tilde \nu}}{x^{\beta}}_{,{\tilde \mu}} {x^{\lambda}}_{,{\tilde \nu}}=0 \, .
\label{eom-x}
\ee
In a $(N+1)$-dimensional flat Friedmann-Robertson-Walker Universe, the line element is given by
\be
ds^2=a^2(\eta)\left(d^2\eta-{d\mathbf{x}}\cdot d{\mathbf{x}}\right)\,,
\ee
where $a$ is the scale factor, $\eta=\int dt/a$ is the conformal time, $t$ is the physical time and $\mathbf{x}$ is a N-vector whose components are comoving cartesian coordinates. Since the Nambu-Goto action is invariant under worldsheet parameterizations, one is free to impose temporal transverse gauge conditions
\bq
u^0&=&\eta\,, \\
\dot{\mathbf{x}}\cdot\mathbf{x}_{,{\tilde i}}&=&0\,. \label{gauge1}
\eq
A dot represents a derivative with respect to conformal time and $\mathbf{x}_{,{\tilde i}} = \partial \mathbf{x}/ \partial u_{\tilde i}$. 
These gauge conditions are chosen so that the timelike worldsheet coordinate is identified with the conformal time and $\dot{\mathbf{x}}$ represents the physical velocity, perpendicular to the $p$-brane itself. Moreover, the local orthogonal coordinate system is chosen in such a way that the coordinate lines coincide with the principal directions of curvature. Eq. (\ref{eom-x}) yields
\bq
\ddot{\mathbf{x}} & + & \left(p+1\right) \mathcal{H}(1-\dot{\mathbf{x}}^2)\dot{\mathbf{x}} =\nonumber\\
& =& \epsilon^{-1} \sum_{{\tilde i}=1}^p\left[\frac{\mathbf{x}_{,{\tilde i}}}{\epsilon}\Pi_{{\tilde j}\neq {\tilde i}}(\mathbf{x}_{,{\tilde j}})^2\right]_{,{\tilde i}}\,,\label{ngeq}\\
\dot{\epsilon} & = & -\left(p+1\right)\mathcal{H}\epsilon \dot{\mathbf{x}}^2\,,
\eq
where
\be
\epsilon=\left(\frac{(\mathbf{x}_{,1})^2\cdots (\mathbf{x}_{,p})^2}{1-\dot{\mathbf{x}}^2}\right)^{\frac{1}{2}} \,,
\ee
and $\mathcal{H}=\dot{a}/a$.

Let us define the unitary vectors
\be
\hat{\mathbf{v}}=\frac{\dot{\mathbf{x}}}{v}\,, \qquad  {\hat{\mathbf{e}}}_{\tilde i}=\frac{\mathbf{x}_{,{\tilde i}}}{\left|{\mathbf{x}}_{,{\tilde i}}\right|}\,,
\ee
where $v=\left|\dot{\mathbf{x}}\right|$ is the velocity of the brane at a particular point and ${\hat{\mathbf{e}}}_{\tilde i}$ (with ${\tilde i}=1,...,p$) are the unitary tangent vectors along the principal directions of curvature. Differentiating Eq. (\ref{gauge1}) with respect to conformal time one finds
\be
{\bf a}_{\tilde i}=\left(\ddot{\bf x}\cdot\hat{\bf e}_{\tilde i}\right)\hat{\bf e}_{\tilde i}=-\frac{v}{\left|{\bf x}_{,{\tilde i}}\right|}v_{,{\tilde i}}\hat{\bf e}_{\tilde i}\,,
\ee
and, consequently, there is a component of the acceleration parallel to the $p$-brane
\be
{\ddot{\mathbf{x}}}_{=} =  \sum_{i=1}^p {\mathbf{a}}_{\tilde i}\,.
\ee
Using Eq. (\ref{ngeq}) one may show that
\be
\ddot{\bf x}-\ddot{\bf x}_{=}=-(p+1)\mathcal{H}(1-v^2)v\hat{\bf v}+\frac{1}{\gamma^2}\sum_{{\tilde i}=1}^{p}{\bf k}_{\tilde i}\,,
\label{acc1}
\ee
where we introduced the comoving curvature vector along the principal direction of curvature ${\tilde i}$ defined by
\be
{\bf k}_{\tilde i}=\left.\frac{\partial \hat{\bf e}_i}{\partial s_{\tilde i}}\right|_{u_{\tilde j}={\rm constant}}\,,
\ee
with ${\tilde j} \neq {\tilde i}$.
Here we introduced the physical length along the ${\tilde i}$ direction $ds_{\tilde i}=\left|{\bf x}_{,\tilde i}\right|du_{\tilde i}$. Using Eq. (\ref{acc1}), taking into account that ${\ddot{\mathbf{x}}}=\dot{v}\hat{\mathbf{v}}+v\dot{\hat{\mathbf{v}}}$ and that $\dot{\hat{\mathbf{v}}}$ is perpendicular to $\hat{\mathbf{v}}$, one finds that the tangential acceleration (parallel to the velocity) is given by
\bq
{\bf a}_{p+1}&=&\ddot{\bf x}_{\parallel}=\left(\ddot{\bf x}\cdot\hat{\bf v}\right)\hat{\bf v}=\nonumber\\
&=&\hat{\bf v}(1-v^2)\left[\sum_{{\tilde i}=1}^p k_{{\tilde i}\parallel}-(p+1)\mathcal{H}v\right]\,,
\eq
where we defined the tangential curvature as the projection of the comoving curvature vectors along the velocity direction $k_{{\tilde i}\parallel}={\bf k}_{\tilde i}\cdot\hat{\bf v}$. This tangential acceleration allows us to obtain an evolution equation for the velocity of the $p$-brane
\be
\dot{v}+\left(1-v^2\right)\left[(p+1)\mathcal{H}v-\kappa_{\parallel}\right]=0\,,
\ee
where the total tangential curvature, $\kappa_{\parallel}=\sum_{{\tilde i}=1}^p k_{{\tilde i}\parallel}$, was introduced. This equation is identical to that obtained from field theory equations in Ref. \cite{Sousa:2011ew}.

There are $N-p-1$ directions which are simultaneously perpendicular to the $p$-brane and to its velocity. Let us denote the unitary vectors along these directions by ${\hat{\mathbf{e}}}_{l}$ with $l=p+2,...,N$. The acceleration along these directions is given by
\bq
{\bf a}_l&=&\left(\ddot{\bf x}\cdot \hat{\bf e}_l\right)\hat{\bf e}_l=\hat{\bf e}_l (1-v^2)\sum_{{\tilde i}=1}^{p}\hat{\bf k}_{\tilde i}\cdot\hat{\bf e}_l= \nonumber \\
&=&\hat{\bf e}_l(1-v^2)\kappa_{\perp l}\,,
\eq
where we have introduced the total comoving curvature along the perpendicular direction $l$ $\kappa_{\perp l}=\sum_{\tilde i=1}^p\hat{\bf k}_{\tilde i}\cdot\hat{\bf e}_l$. Therefore, the total perpendicular acceleration is given by
\be
\ddot{\bf x}_{\perp}=(1-v^2)\sum_{l=p+2}^{N}\hat{\bf e}_l\kappa_{\perp l}\,,
\ee
with ${\ddot{\mathbf{x}}}={\ddot{\mathbf{x}}}_{=}+{\ddot{\mathbf{x}}}_{\parallel}+{\ddot{\mathbf{x}}}_{\perp}$.

\section{VOS model\label{VOS}\label{vos-sec}}

Let us consider a network of $p$-branes in a $(N+1)$-dimensional FRW Universe and define the root mean square (RMS) velocity, ${\bar v} ={\sqrt {\langle v^2\rangle}}$ as
\be
{\bar v}^2 = \frac{\int v^2 \epsilon d^p u}{\int \epsilon d^p u}\,,
\ee
where $d^p u=du^1 \times \,...\, \times du^p$. The characteristic length, $L$, of the network is defined as
\be
{\bar \rho} = \frac{\sigma_p}{L^{N-p}}\,,
\ee
where ${\bar \rho} $ is the average brane density. 

A unified VOS model for the dynamics of $p$-brane networks in $(N+1)$-dimensional FRW universes was derived in \cite{Sousa:2011ew}. This model is described by the following equations
\bq
\frac{d{\bar v}}{dt}&+&\left(1-{\bar v}^2\right)\left[\frac{{\bar v}}{\ell_d}-\frac{k}{L}\right]\,, \label{VOS_v}\\
\frac{dL}{dt}&=&HL+\frac{L}{D \ell_d}{\bar v}^2+ \frac{{\tilde c}}{D}{\bar v}\,,
\label{vos-L}
\eq
where $\ell_d^{-1}=(p+1)H$ is the damping lengthscale, $H={\mathcal H}a$ is the Hubble parameter, $D=N-p$, ${\tilde c} \ge 0$ is the energy-loss parameter, $k={\bar \kappa}_{\parallel}L/a$ is a dimensionless curvature parameter, 
\be
{\bar \kappa}_{\parallel}=\frac{\left<v\left(1-v^2\right) \kappa_{\parallel} \right>}{{\bar v}\left(1-{\bar v}^2\right)}=\frac{\int v\left(1-v^2\right)\kappa_{\parallel} \epsilon d^p u}{{\bar v}\left(1-v^2\right)\int \epsilon d^p u}
\ee
and the assumption that $\left<v^4\right>={\bar v}^4$ was made (see \cite{Martins:1996jp}). A frictional force --- caused by the interaction of the branes with ultrarelativistic particles or other frictional sources --- may be included in Eq. (\ref{VOS_v}), by introducing an extra term in the damping lenghtscale, $\ell_d^{-1}=(p+1)H+\ell_f^{-1}$.

Consider the presence of an interaction mechanism between the $p$-branes and a component average density $\rho_{int}$. A very conservative upper limit to the total momentum per unit volume transfered from that component to the $p$-branes in one Hubble time,
\be
\left|\frac{d{\bf p}}{dV}\right|\sim \frac{\bar \rho}{H} \frac{dv}{dt}\,,
\ee
is given by $\rho_{int}$. In this case one has
\be
\left|\frac{d{\bar v}}{dt}\right|_{int} \lsim \chi H\,,
\ee
where $\chi=\rho_{int}/\bar \rho$. 

If $a\propto t^\beta$ (with $0<\beta<1$), and the friction scale is negligible compared to the Hubble radius ($\ell_f \ll H^{-1}$), this model may admit linear scaling solutions of the form
\be
L=\xi t \qquad \mbox{and} \qquad \bar v=\mbox{constant}\,.
\label{linearscalinga}
\ee
See \cite{Sousa:2011ew} for a discussion of other scaling regimes.

If $\chi={\rm constant}$, then
\bq
\xi &=& \sqrt{\left|\frac{k(k+{\tilde c})}{\beta (1-\beta)D(p+\chi+1)}\right|}\,, \label{linearscaling1}\\ 
{\bar v} &=& \sqrt{\frac{(1-\beta)kD}{\beta(k+{\tilde c})(p+\chi+1)}}\,.\label{linearscaling2}
\eq
The effects of this interaction mechanism may slow down the branes slightly but, if the interacting component of energy density $\rho_{int}$  is subdominant (that is $\chi \lsim 1$), its potential role on the frustration of the network (defined by $L\ll H^{-1}$ and $v\ll 1$) is very limited. Relaxing the assumption that the $p$-brane network is the dominant energy component may help frustration:  if $\chi\gg 1$ (or equivalently if $\rho_{int}\gg {\bar \rho}$) the interaction mechanism may decelerate the branes effectively, leading to the frustration of the network. However, in this case, $\rho_{int}$ would be the main contributor to the energy budget. Of course, assuming that $\chi$ is time independent is unrealistic, since one would expect the expansion of the background to affect the efficiency of any realistic interaction mechanism (in particular, the ratio $\chi=\rho_{int}/\bar \rho$ is expected to be a function of $a$). Nonetheless, Eq.  (\ref{VOS_v}) shows that considering a time varying $\chi$ does not help much if its present value $\chi_0 \lsim 1$. As a matter of fact, frustration might only occur, under these circumstances, for networks which have $k\ll 1$ for $v\ll 1$. This generalizes a well known result for domain walls \cite{PinaAvelino:2006ia,Avelino:2008ve,Sousa:2009is} to arbitrary $p$-brane networks. In the particular case of domain walls \cite{Avelino:2008ve}, there is very strong analytical and numerical evidence that  domain wall networks (with or without junctions) are unlikely to attain $k \ll 1$ for $v\ll 1$, if they are the dominant energy component. This effectively rules out domain walls as a cosmologically relevant dark energy candidate: frustration can only occur either if the network is designed to have $k\ll 1$ in the non-relativistic limit --- which appears to be unrealistic --- or if $\chi$ is much larger than unity --- in which case the domain wall energy density would be subdominant. For $p<N-1$, the damping is less efficient, and thus frustration is even less likely to result from the natural evolution of the network. Therefore, unless there was a natural mechanism that drives $k$ towards zero in the non-relativistic limit (which seems unlikely) the "no-frustration conjecture" is also expected to apply to any realistic and cosmologically relevant $p$-brane network.

In ref. \cite{Battye:2010dk} the authors perform field theory simulations of a model with $\mathbb{Z}_2\times U(1)$ symmetry in (2+1)-dimensions. Their  model has two discrete vacua, allowing for domain walls and a conserved Noether charge. The authors argue that the Noether charge and currents become localized on the walls, forming kinky vortons and providing a possible mechanism for the frustration of domain wall networks. However, the authors never calculate the overall equation of state of the network. Had they done that, they would have found significant deviations with respect to that of a frustrated featureless domain wall gas ($w=-2/3$).

\section{Conclusions \label{conc}}

In this paper, we derived the equation of motion for infinitely thin featureless $p$-branes, by computing the tangential and normal components of the acceleration directly from the Nambu-Goto action. Our results further validate the semi-analytical VOS model developed in \cite{Sousa:2011ew}, and its use in dynamical studies of $p$-brane networks. The VOS model unifies, in a single framework, the dynamics of $p$-brane networks for any possible values of the pair $(N,p)$. While part of the dynamical dependency on the parameters $N$ and $p$ is explicit in the equations of motion, there is also an implicit dependency in the parameters ${\tilde c}$, $\ell_f$ and $k$. In this paper we demonstrated that, if the $p$-branes are the dominant component of the universe, then frustration is not possible except if the curvature parameter is driven towards very small values for non-relativistic networks or if the expansion is accelerated. In the case of domain walls there is very strong analytical and numerical evidence (both in two $(N=2,p=1)$ and three $(N=3,p=2)$ spatial dimensions) that $k$ is never becomes much smaller than unity (except deep into inflationary or friction dominated regimes), thus preventing frustration from being attained, at least for realistic domain wall networks playing a dark energy role.  We conjecture that this may be a general result, valid for any realistic $p$-brane network independently of the values of $N$ and $p$ with $1 \le p \le N-1$.

\begin{acknowledgments}

This work is partially supported by FCT-Portugal through project CERN/FP/116358/2010.

\end{acknowledgments}


\bibliography{dw}

\end{document}